\documentclass[aps,prl,twocolumn,groupedaddress,showpacs]{revtex4}
\usepackage{epsfig}
\usepackage{graphicx}
\usepackage{amsfonts}
\usepackage{latexsym}
\usepackage{bm}
\begin{document}

%\preprint{}

\title{Anderson localization from the replica formalism}
\author{Alexander Altland$^1$, Alex Kamenev$^{2}$, and Chushun Tian$^{2,3}$}
%\thanks{}
\affiliation{$^1$ Institut f{\"u}r Theoretische Physik,
Universt{\"a}t zu K{\"o}ln, K{\"o}ln, 50937, Germany\\
$^2$ Department of Phhysics, University of Minnesota, Minneapolis
MN 55455, USA\\
$^3$ Kavli Institute for Theoretical Physics, University of
California, Santa Barbara CA, 93106, USA }


\date{\today}

\begin{abstract}
We study  Anderson localization  in quasi--one--dimensional
disordered wires within the framework of the replica
$\sigma$--model. Applying a semiclassical approach (geodesic
action plus Gaussian fluctuations) recently introduced within the
context of supersymmetry by Lamacraft, Simons and Zirnbauer
\cite{LSZ}, we compute the {\em exact} density of transmission matrix
eigenvalues of superconducting wires 
(of symmetry class
$C$I.) For the unitary class 
of metallic systems (class $A$) we are able to obtain the density function, save for its
large transmission tail.
\end{abstract}


\maketitle

At present, there exist two theoretical approaches capable of
describing strongly localized phases of disordered wires:
supersymmetry (SUSY) \cite{Efetov} and the DMPK transfer matrix
approach \cite{DMPK}. This represents a serious limitation in as much
as both formalisms are ill--suited for generalization to the presence
of Coulomb interactions (see, however, Ref.[\onlinecite{Efetov04}].)
Reciprocally, it has, so far, not been possible to describe strong
localization phenomena by those theories that may be applied to the
analysis of interaction effects --- replica field
theory~\cite{Finkelshtein} and the Keldysh approach~\cite{Keldysh}.

It is the purpose of this letter to introduce a replica field theory
approach, capable of describing strongly localized phases.
Conceptually, our work is based on a recent paper \cite{LSZ} by
Lamacraft, Simons and Zirnbauer (LSZ) where saddle--point techniques
have been applied to analyze the SUSY generating functionals of quasi
one--dimensional disordered conductors.  Specifically it was shown
that four out of ten symmetry classes of disordered metals are
semiclassically exact \cite{footnotesp} in that the stationary phase
results coincide with those obtained by DMPK methods~\cite{Brouwer02}.
We here show that the phenomenon of semiclassical exactness pertains
to the replica formalism and, in particular, `survives' the analytical
continuation inherent to that approach. Applying the technique to the
non--semiclassically exact unitary symmetry class, we find that it
still produces qualitatively correct results.

To introduce the replica--generalization of the method we consider a
disordered superconducting wire in the presence of spin--rotation and
time reversal invariance (symmetry class $C$I in the classification of
Ref.~[\onlinecite{AZ97}].) The (thermal) transport properties of this
system may be conveniently characterized in terms of the average
density of transmission matrix eigenvalues, $\rho(\phi)$. Within the
fermion--replica formalism the latter may be expressed through the
generating function
$$
{\cal Z}(\hat \theta)\equiv \prod_{a=1}^R \det
  \left(1-\sin^2(\theta_a/2)\,{\bf t}{\bf t}^\dagger \right),
$$
where ${\bf t}{\bf t}^\dagger$ is the transmission matrix with
eigenvalues ${\cal T}_j = \cosh^{-2}(\phi_j / 2)$ and $\hat\theta
\equiv{\rm diag}(\theta_1,\dots, \theta_R)$. Defining the function
$F(\theta)\equiv \lim_{R\to 0} {d\over
d\theta_1}\big|_{\theta_a\to \theta}{\cal Z}(\hat\theta)$, the
transmission matrix eigenvalue density is obtained
as~\cite{Nazarov}: $\rho(\phi)= {1\over 2\pi} (F(i\phi + \pi) -
F(i\phi - \pi))$.

The field theoretical representation of the generating function for
class $C$I is given by
\begin{equation}
  \label{eq:2}
  {\cal Z}(\hat\theta) = \hspace{-.2cm}
  \int\limits_{g(0)}^{ g(T)}
  \hspace{-.15cm}\!{\cal D}g\,\,e^{-S[g]}, \qquad S[g]={1\over
  8}\int\limits_0^T\!\! dt\,\, {\rm tr}(\partial g \partial
  g^{-1})\,,
\end{equation}
where $g$ is a field of matrices $g(t)\in {\rm Sp}(2R)$, the
functional integration extends over the Haar measure on the
symplectic group, and $T=L/\xi$ is the length of the wire, $L$, in
units of the localization length $\xi$.  At the left and right
end point of the wire the field is subject to boundary
conditions~\cite{Nazarov,Rejaei96} which in the case of class
$C$I read as $g(0)=\openone$ and $g(T) = \exp(i\hat\theta \otimes
\sigma_3)$. Here the Pauli matrix $\sigma_3$ acts in the space
defining the symplectic condition $g^{-1}=\sigma_2 g^T \sigma_2$.

Our strategy will be to subject the functional (\ref{eq:2}) to a
straightforward stationary phase analysis~\cite{footnotecon}.
Varying the action $S[g]$ w.r.t. $g$, one  obtains the
Euler--Lagrange equation: $\delta_g\big|_{g=\bar g} S[g]=0
\Rightarrow \partial (\bar g^{-1}
\partial \bar g)=0$, which integrates to the condition $\bar
g^{-1} \partial \bar g={\rm const.}$ The solutions to this latter
equation are given by $\bar g = \exp(i \bar W t/T)$, with constant
Lie--algebra elements $\bar W\in {\rm sp}(2R)$. Evaluating $\bar
g$ at the system boundary $t=T$, we obtain the condition $\exp(i
\bar W) = \exp(i \hat\theta \otimes \sigma_3)$. This is solved by
$\bar W \equiv \hat \theta^{(n)}\otimes \sigma_3$, where $\hat
\theta^{(n)}\equiv \hat\theta + 2\pi \hat n$ and $\hat n={\rm
diag}(n_1,\dots,n_R)$  is a vector of integer `winding numbers'.
% Geometrically, each configuration $\bar g\equiv a^{(n)} \equiv
% \exp(i W^{(n)} t/T)$ parameterizes a geodesic trajectory on the group
% manifold connecting the origin $g=\openone$ with the final
% configuration $g=a(\delta)$. Mean field configurations with
% non--vanishing winding numbers spin repeatedly around the `torus'
% $T\subset {\rm Sp}(2R)$ of the group, i.e. its maximal commutative
% subgroup (aka the group of diagonal matrices.)
The saddle point action is given by
$S[\bar g^{(n)}]={1\over 4T}\sum_{a=1}^R (\theta^{(n)}_a)^2$,
indicating that at
length scales, $T\gtrsim 1$, mean field configurations
traversing multiply around the group manifold become energetically
affordable. Physically, these configurations describe the massive
(and perturbatively inaccessible) buildup of interfering
superconductor diffusion modes. Their proliferation at large
length scales forms the basis of the localization phenomenon.

To obtain the contributions of individual saddle points, $\bar
g^{(n)}$, to the generating function, we need to integrate over
quadratic fluctuations. We thus generalize to field configurations
$g(t)= \exp(iW(t))\, \bar g^{(n)}$, where the fields $W(t)\in {\rm
sp}(2R)$ obey vanishing (Dirichlet) boundary conditions
$W(0)=W(T)=0$. Parameterizing  these fields as $W=\sum_{\mu=0}^4
W_\mu \otimes\sigma_\mu$, where $\sigma_0=\openone_2$ and $W_\mu$
are $R\times R$ hermitian matrices subject to the  Lie algebra
constraints $W_0=-W_0^T$ and $W_i=W_i^T$, $i=1,2,3$, the quadratic
expansion of the action reads as: $S[g] = S[\bar g^{(n)}] +
S_I[W_0,W_3] + S_{II}[W_1,W_2] + {\cal O}(W^3)$, where
%\begin{widetext}
  \begin{eqnarray*}
    S_I[W_0,W_3]&=& {1\over 4} \int\limits_0^T\!\! dt\, {\rm tr}\Big(\partial W_0 \partial
  W_0 + \partial W_3 \partial W_3\\
 &-& {2i\over T} (W_0\partial W_3 +
  W_3 \partial W_0)\, \hat\theta^{(n)} \Big),\\
 S_{II}[W_1,W_2]\!&=&\! {1\over 2} \int\limits_0^T\!\! dt\, {\rm
tr}\left(\partial W_1 \partial
  W_2  + i { \epsilon_{ij3}\over T} W_i\partial W_j\,  \hat\theta^{(n)}\right).
    \end{eqnarray*}
%\end{widetext}
    The integration over the matrices $W_\mu$ leads to fluctuation
    determinants, which may be calculated by the auxiliary identity
    $\det(-\partial_t^2+ 2 z T^{-1} \partial_t)= \sinh(z)/z$, where
    $z\in \Bbb{C}$, and the differential operator acts in the space of
    functions obeying  Dirichlet boundary conditions. As a result
    we obtain the stationary phase generating function
\begin{widetext}
 \begin{equation}
  \label{eq:3}
  {\cal Z}(\hat\theta )=\sum\limits_{\{n\}} \prod\limits_{a<a'}^R {\left(\theta^{(n)}_a -
  \theta^{(n)}_{a'}\right)/2\over \sin\left[(\theta^{(n)}_a -
  \theta^{(n)}_{a'})/2\right]}\, \prod\limits_{a\le a'}^R {\left(\theta^{(n)}_a +
  \theta^{(n)}_{a'}\right)/2\over \sin\left[(\theta^{(n)}_a +
  \theta^{(n)}_{a'})/2\right]}\,\,\exp\left(-{1\over 4T}\sum_{a=1}^R
  (\theta^{(n)}_a)^2\right),
\end{equation}
\end{widetext}
where the first/second fluctuation factor stems from the integration
over the field--doublets $(W_0,W_3)$/$(W_1,W_2)$. (In passing, we note
that as an alternative to the brute force integration outlined above
the result (\ref{eq:3}) can be obtained by group theoretical
reasoning: according to general principles \cite{Picken}, the
fluctuation integral around extremal (geodesic) configurations $\bar
g^{(n)}$ on a general semi--simple Lie group is given by: $
\prod_{\alpha>0} {\alpha(\bar g^{(n)})\sin^{-1}(\alpha(\bar g
  ^{(n)}))} \exp(-S[\bar g ^{(n)}])$, where the product extends over
the system of positive roots of the group, $\alpha (\bar g^{(n)})$.
Equation~(\ref{eq:3}) above is but the ${\rm Sp}(2R)$--variant of this
formula.)

%% At this stage, a few remarks are in order:
%% (i) As an efficient alternative to the direct integration procedure
%% outlined above, the result (\ref{eq:3}) may be obtained by
%% group--theoretical reasoning: according to general principles[], the
%% integral over fluctuations around a trajectory $(t/T) \theta^{(n)}$ on
%% the torus of a general semi--simple Lie group is given by
%% $\sum_{\{n\}} \prod_{\alpha>0} {\alpha(\theta^{(n)})\over
%%   2\sin(\alpha(\theta_n))} \exp(-S[\theta^{(n)}])$, where the product
%% extends over the system of positive roots[] of the group.
%% Eq.(\ref{eq:3}) above is but the ${\rm Sp}(2R)$--variant of this
%% formula. (ii) A variant of the Poisson sum rule\footnote{See [Picken]
%%   for the general mathematics of this transformation and [dowker] for
%%   an application to the group ${\rm SU}(N)$.} may be applied to map
%% the sum over winding numbers onto the spectral decomposition of the
%% partition function discussed above. The weights $\lambda$ are `Fourier
%% conjugate' to the winding numbers very much like, say, momentum
%% quantum numbers of a particle on a ring are Fourier conjugate to real
%% space winding numbers.

In the limit of coinciding boundary phases, $\theta_a \to \theta$,
the denominators $\sin[(\theta^{(n)}_a - \theta^{(n)}_{a'})/2]\to
0$, i.e. the contribution of configurations $\hat n$ containing
non--vanishing winding number differences $n_a - n_{a'}\not=0$
 diverges. (At the same time, we do know that the
integration over the {\it full} group manifold must generate a
finite result. Indeed, it turns out that if we first sum over all
winding number configurations $\hat n$ and only then take the
limit of coinciding phases, all divergent factors disappear.) This
divergence reflects the presence of a zero mode in the system: for
uniform boundary phases, $\hat \theta \propto \openone_R$,
transformations $\bar g^{(n)} \to \exp(i V^0)\bar g^{(n)}
\exp(-iV^0)$ with constant block--diagonal $V^0=V_0^0
\otimes\sigma_0 + V^0_3 \otimes\sigma_3$ conform with the boundary
conditions but do not alter the action.

%To better understand the significance of these zero modes,
%consider the particular winding number configuration

As we shall see below, the presence of zero modes implies that
only winding number configurations of the special form
$(n,0,\dots,0)$ survive the replica limit, $R\to 0$. However,
before elaborating on this point, let us evaluate the contribution
${\cal Z}_n$ of the distinguished configurations to the generating
function. Throughout we will denote the boundary angles by
$\theta_a\equiv \theta +\eta_a$, understanding that the limit
$\eta_a\to 0$ is to be taken at some stage. (Within this
representation, the `free energy' $F(\theta)\partial_{\theta_1}\big|_{\theta_a \to \theta} {\cal Z}(\hat
\theta)= \partial_{\eta_1}\big|_{\eta_a\to 0} {\cal Z}(\theta+\hat
\eta)$.) The `dangerous' product $\prod_{a<a'}(\dots)$ in
Eq.~(\ref{eq:3}) then reduces to $ \sim (\pi
n/\sin(\eta_1/2))^{R-1}\approx (2\pi n/\eta_1)^{R-1}$; all other
contributions to ${\cal Z}_n$ are finite. The appearance of a pole
of $(R-1)$st order hints at the presence of $R-1$ complex zero
modes (generated by the $R-1$ components of the matrix $V^0$ that
do not commute with $\hat g^{(n)}$.) At this stage, we take the
limit $R\to 0$.
%In concrete terms, this amounts to setting $R=0$ in all regular
%contributions to ${\cal Z}_n$, while
As a result, the divergent factor gets replaced by a `pole of
degree $(-1)$', i.e. the zero: $\eta_1/ (2\pi n)$. (It is
worth  noting that in SUSY a contribution similar to the
singularity of degree $(-1)$ is obtained by integration over the
non--compact bosonic degrees of freedom; the complementary single
replica channel $a=1$ corresponds to the fermionic sector.)
Therefore  the subsequent differentiation ($F[\phi]\sim
\partial_{\eta_1}\big|_{\eta_a\to 0}{\cal Z}$) {\em must} act
on this linear factor $\eta_1$, all other occurrences of $\eta_a$
in ${\cal Z}_n$ may be ignored.

\begin{figure}
\includegraphics[width=8cm]{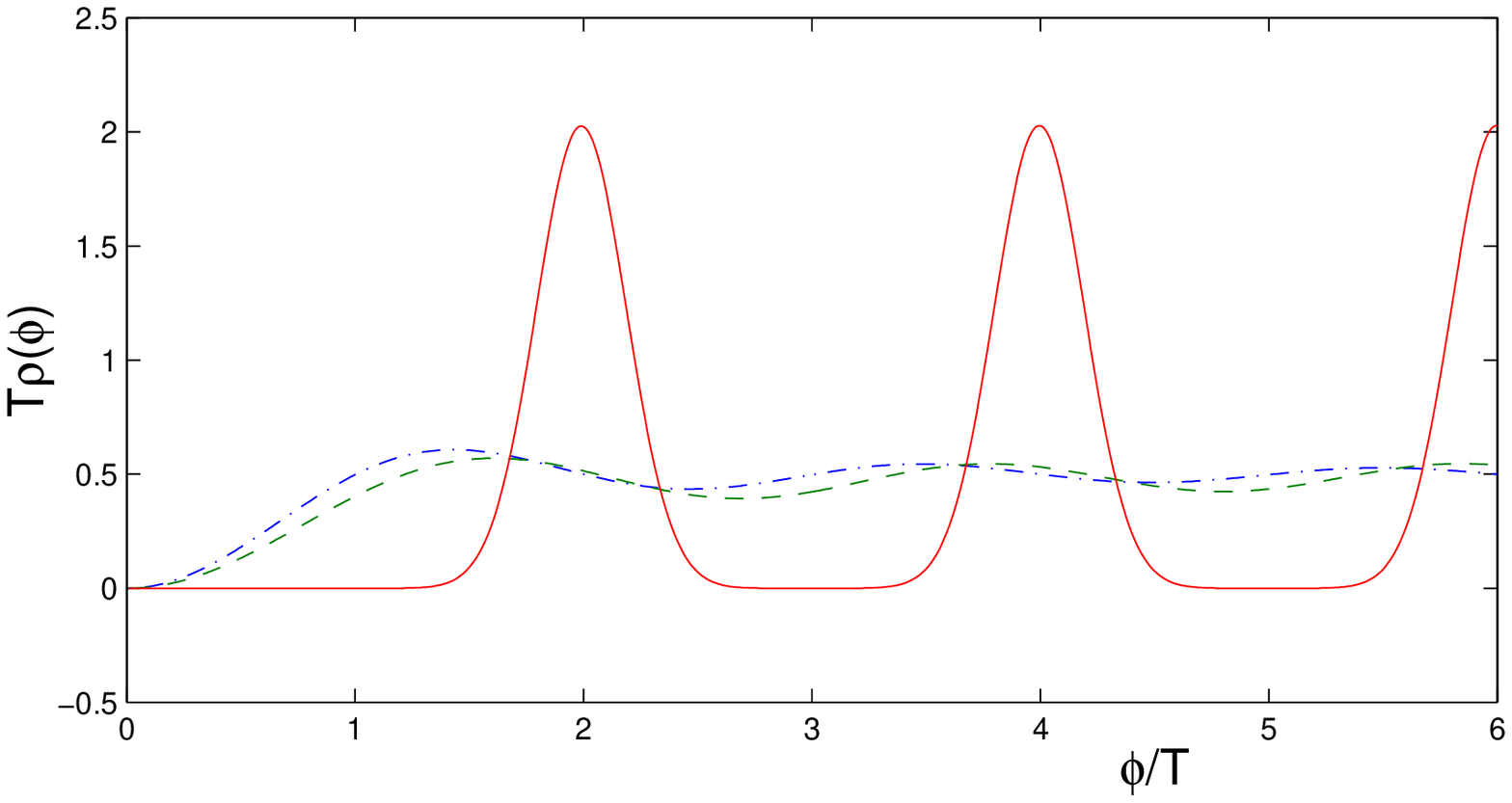}
\includegraphics[width=8cm]{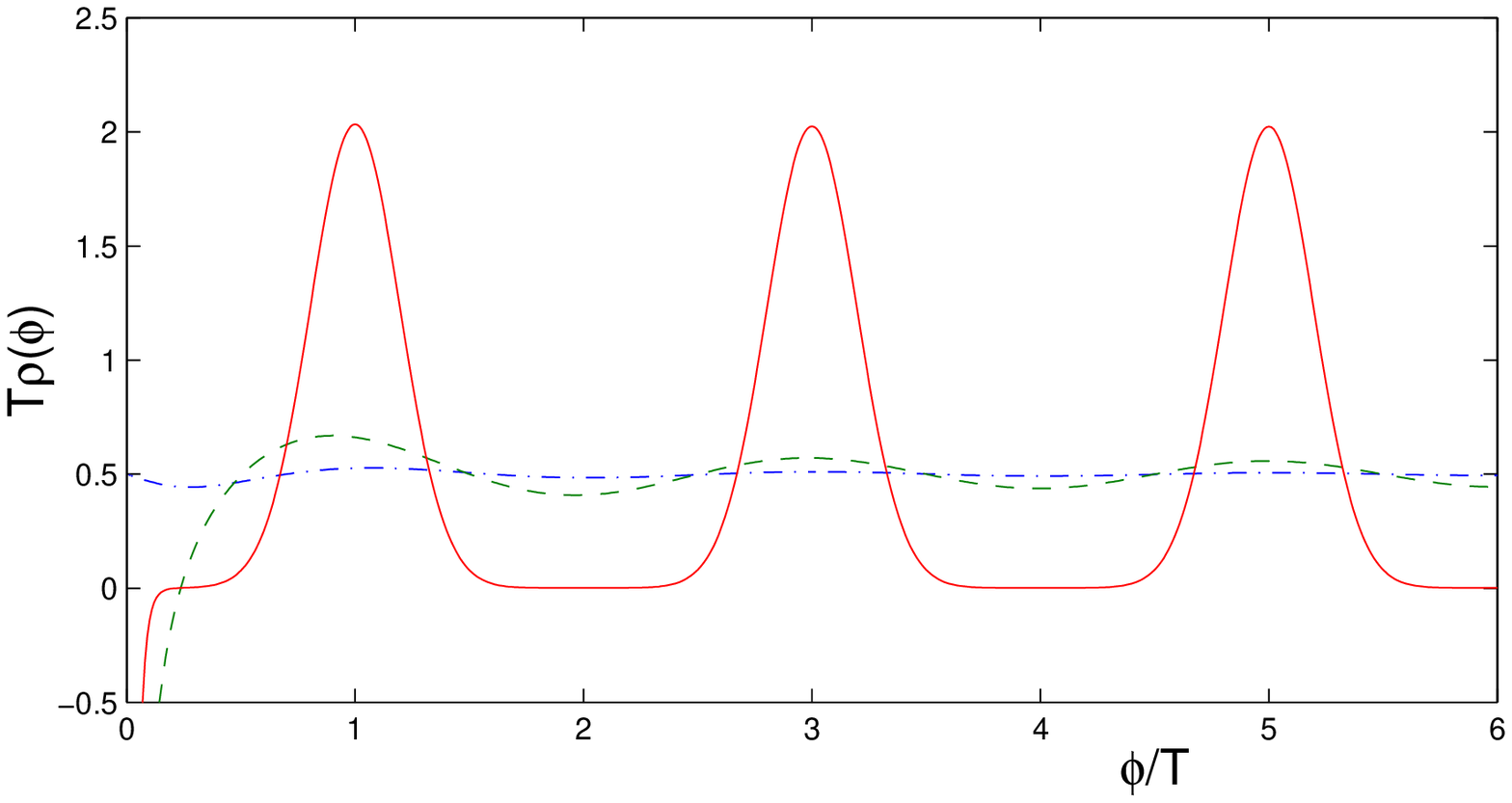}
\caption{Top: Density of transmission eigenvalues of the
  superconductor class $C$I for T=.02 (dashed--dotted); 1 (dashed); 50
  (full). Bottom: the same for the unitary class $A$. The negative
  density at small $\phi$ represents an artefact of the
  saddle--point approximation. }
\label{fig1} \vskip -.5cm
\end{figure}

Evaluating the partition function in this manner, we obtain $ {\cal
  Z}_{n\not=0} = {\eta_1\over 2\pi n} {\theta+2\pi n\over \theta+\pi
  n} \exp(-\pi n(\pi n + \theta)/T)$. We finally differentiate w.r.t.
$\eta_1$ and arrive at the result $\rho(\phi)=\rho_0(\phi) +
\sum_{n\not=0} \rho_n(\phi)$, where the `Drude plus weak localization
term' $\rho_0= (2T)^{-1} -(\phi^2+\pi^2)^{-1}/2$,
while the non--perturbative contributions are given by:
\begin{equation}
                                             \label{rhoCI}
\rho_n(\phi)= - {e^{-{\pi^2 \over T} n ( n+1)}\over
  2\pi^2 n}\,{\rm Re}\left[{\phi + i \pi (2n+1)\over \phi + i \pi(n+1)}\,
  \, e^{i {\pi n \phi \over T}}\right].
\end{equation}
This expression identically coincides with the SUSY
result~\cite{LSZ}, and with the exact DMPK
result~\cite{Brouwer02}. To illustrate the `crystallization' of
the transmission matrix eigenvalues at the discrete values $\phi_j
\approx 2jT$, the function $\rho(\phi)$ is plotted in
Fig.~\ref{fig1}a for a few values of $T$. Following LSZ, the heat
conductance of the wire may be obtained by integrating the result
above against the weight function $1/\cosh^2(\phi/2)$. Summing the
result of this integration over winding numbers, one obtains the
asymptotic result~\cite{LSZ} 
$g\stackrel{T\gg1}{\simeq} 4\, e^{-T}/\sqrt{\pi T} $ .

Our so far analysis focused on the specific set of winding
number configurations, $(n,0\dots,0)$. To understand why
contributions of different structure vanish  --- a fact that
greatly simplifies the formalism ---
consider the set $(0,\dots,n,\dots,0)$. By symmetry,
winding number configurations of this type will lead to an
expression similar to ${\cal Z}_n$ above, only that the leading
pre--factor gets replaced: $\eta_1/(2\pi n)\to \eta_a/(2\pi n)$,
where $a\in \{2,\dots,R\}$ marks the position of the
non--vanishing winding number. Since, however, we still
differentiate w.r.t. $\eta_1$, this contribution vanishes in the
limit $\eta_a\to 0$. The argument above may be generalized to
generic contributions, $(n_1,n_2,\dots,n_R)\not=(n,0,\dots,0)$.
(By symmetry, one may order the winding numbers in an ascending
order $(0,\dots,0,1,\dots,1,2\,\dots)$. Assuming that there are
$N_n$ winding numbers $n$ (where $\sum_n N_n =R$) and choosing the
boundary angle in the sector $n$ to be $\theta+ n \eta$, one
verifies that for any fixed configuration, the $R\to 0$ result
contains uncompensated powers of $\eta$ and, therefore, vanishes.)

Before proceeding, it is worthwhile to compare the mean field
analysis above to the more established field theory transfer
matrix method~\cite{Efetov}. To this end, let us interpret ${\cal
Z}(\hat\theta)=\langle
 g(T) |\exp(- T \hat H)|\openone\rangle$ as the path integral
 describing the (imaginary time) quantum mechanical transition amplitude
 $|\openone \rangle \stackrel{T}{\to} |g(T)\rangle$ of a particle on
 the group space ${\rm Sp}(2R)$. The Hamiltonian corresponding to the
 (purely `kinetic') action of the path integral is given by $\hat H
 =-2\Delta$ where $\Delta$ is the Laplace operator of the
 group space ${\rm Sp}(2R)$.
 
 Our analysis above has been tantamount to a semiclassical or WKB
 analysis of the transition amplitude.  Alternatively, and more
 rigorously, one may employ the spectral decomposition,
 ${\cal Z}(\hat \theta)=\sum_\lambda \psi_\lambda^\ast(g)
 \psi_\lambda(\openone) \, \exp(- T \epsilon_\lambda)$, where
 $\psi_\lambda$ are the eigenfunctions of the Laplace operator,
 $\epsilon_\lambda$ its discrete energy eigenvalues and $g\equiv g(T)$. For general Lie
 groups (and supergroups) formal expressions for these spectral
 decompositions are known~\cite{footnoteweight}. Noting that for large
 systems $L\gg \xi$, only eigenstates with minimal energy
 $\epsilon_\lambda$ effectively contribute to the sum, this knowledge
 has been used to compute the localization properties of disordered
 quantum wires within the SUSY formalism~\cite{Efetov,Rejaei96}. The
 problems with transferring this approach to the replica formalism lie
 with the analytical continuation from integer group dimension $R$ to
 $R\to 0$.  In taking this limit, it is essential to keep track of
 high--lying contributions to the spectral sum. These
 terms grow rapidly more complex which is why attempts to obtain a
 replica variant of the `quantum approach' above have failed so far.

 Having discussed the method for a symmetry class that enjoys the
 semiclassical exactness, we next outline what happens in cases where
 this feature is absent. By way of example, consider a metallic
 disordered quantum wire in the absence of time--reversal invariance
 --- the unitary symmetry class, $A$. In this case, the fermionic
 replica generating function is given by ${\cal
   Z}(\hat\theta)=\int_{Q(0)}^{Q(T)} {\cal D}Q\,\exp\left(-{1\over
     8}\int_0^T\!\! dt\,{\rm tr}(\partial Q)^2\right)$, where the
 matrix $Q(t)\in {\rm U}(2R)/{\rm U}(R)\times {\rm
   U}(R)$~\cite{Finkelshtein,KM}, and the boundary configurations are
 given by $Q(0)=\sigma_3\otimes\openone $ and
 $Q(T)=e^{-i\sigma_2\otimes\hat\theta/2} \sigma_3\,
 e^{i\sigma_2\otimes\hat\theta/2}$. Here, the two--component structure
 distinguishes between advanced and retarded indices.  As before, the
 stationary phase configurations: $\bar
 Q(t)=e^{-i\sigma_2\otimes\hat\theta^{(n)}t/(2T)} \sigma_3\,
 e^{i\sigma_2\otimes\hat\theta^{(n)}t/(2T)}$ of the functional
 integral do not mix different replica channels.  Geometrically, they
 can be interpreted as trajectories (in general, with non--zero winding
 number, $n$) on the meridian of the sphere ${\rm U}(2)/{\rm
   U}(1)\times {\rm U}(1)$ (the single replica manifold.)
 Fluctuations may be conveniently parameterized by generalization
 $\sigma_3\to e^{iW(t)} \sigma_3$, where $W=W_1\otimes\sigma_1 +
 W_2\otimes\sigma_2$ and $W_{1,2}$ are hermitian $R\times R$ matrices.

The subsequent calculations largely parallel those for class $C$I
above. Expanding to  second order in $W_{1,2}$ and performing the
Gaussian integration, we again observe that only winding number
configurations $(n,0,\dots,0)$ survive the analytical continuation
procedure, $R\to 0$. Differentiating w.r.t. $\theta_1$ and then
putting $\theta_a\to\theta$, we obtain the result: $\rho(\phi)(2T)^{-1}\! - \sum_{n\not=0} (-1)^n \rho_n(\phi)$, where
$$
\rho_n=  {e^{-{\pi^2n(n+1)\over T} }\over 2\pi^2 n}\, {\rm
Re}\left[{\sqrt{(\phi\!+\!i\pi)
    (\phi\!+\!i\pi(2n\!+\!1))}\over \phi+i\pi(n+1)}\,\, e^{i {\pi n
      \phi\over T}}\right]\, .
$$
(The same result is obtained by  saddle--point analysis of
the SUSY generating functional.)  In Fig.~\ref{fig1}b, the
function $\rho(\phi)$ is plotted for several values of the
parameter $T$. For small $T$ the density is almost constant,
reflecting the Dorokhov distribution of
eigenvalues~\cite{DMPK,Nazarov}.  For large values of $T$ the
spectrum crystallizes at $\phi_j \approx (1+2j)T$.  The lowest
eigenvalue $\phi_0$ does, indeed, correctly determine the
localization length of the system. Except for the evident failure
of the method at small values $\phi\ll \phi_0$ \cite{footnotephi},
the large scale profile of the DoS is in good agreement with
results obtained by the transfer matrix methods
\cite{Rejaei96,Rejaei94,Fraham}.

Summarizing we have shown how the localization phenomenon in quasi
one--dimensional systems may be described by a semiclassical approach
to fermionic--replica field theories. We were able to reproduce the
exact transmission matrix eigenvalue density for symmetry class $C$I,
while for the unitary class we obtain qualitatively correct results
(except for the tails of the eigenvalue spectrum.) The comparative
simplicity of the approach makes us believe that it may be
successfully applied to problems that can not be treated by other
means.  Evidently, the next direction of research will be the study of
the impact of  Coulomb interactions on the localization phenomenon.

We are grateful to M.~R.~Zirnbauer for numerous valuable discussions.
This work is supported by SFB/TR 12 of the Deutsche
Forschungsgemeinschaft (A.~A.), A.~P.~Sloan foundation and the NSF
grants No. DMR--0405212 (A.~K.), DMR--0439026 and PHY-9907949 (C.~T.).


\begin{thebibliography}{}
\bibitem{footnotesp} It is a well--known mathematical fact that
  integrals over certain manifolds (manifolds possessing a symplectic
  structure) may be exactly evaluated by stationary phase methods.
  `Zero--dimensional' applications of this principle include the
  Itzykson--Zuber formulae \cite{IZformula}, and the mean--field
  approach to spectral correlations for certain symmetry classes. By a
  (not rigorously proven) conjecture, semiclassical exactness pertains
  to certain infinite--dimensional manifolds, the so--called loop
  groups \cite{Picken}. From a mathematical point of view, the quantum
  fields describing disordered wires of the four symmetry classes in
  question form manifolds of this type.
\bibitem{footnotecon} On dimensional grounds, the minimal action of a spatially
non--uniform field configuration scales as $S[\bar g]_{\rm
  min}\sim T^{-1}$. To formally justify stationary phase approximation
schemes we must require $T^{-1}>1$\,.
\bibitem{footnoteweight} Referring for a more detailed discussion
  to~\cite{Picken}, we note that for a group valued path integral, the
  spectral sum extends over all irreducible group representations
  while the product of wave functions $\psi_\lambda^\ast(g)
  \psi_\lambda(\openone) = \chi_\lambda(g) \equiv {\rm tr}_\lambda(g)$
  is given by the character $\chi_\lambda$ of $g$ (i.e. the trace of
  the matrix representing the group element $g$ in the representation
  $\lambda$.) The eigenvalue $\epsilon_\lambda = 2 c_2(\lambda)$,
  where $c_2(\lambda)$ is the eigenvalue of the quadratic Casimir
  operator.  (The unique overall commutative operator quadratic in the
  operators representing the Lie algebra in $\lambda$.) The spectral
  sums can be obtained by solving the `Schr\"odinger' equation of the
  problem, or by subjecting winding number sums such as (\ref{eq:3})
  to a Poisson summation scheme.
\bibitem{footnotephi}On physical grounds, one would expect the DoS to
    be exponentially small at $\phi\ll \phi_0$. To understand the
    failure of the semiclassical method, notice that for $\phi\to 0$,
    the mean field configurations are geodesics starting and ending at
    the same point on the sphere. Unlike with generic values
    of the revolution angle, $\theta$, there is a continuous manifold of such
    trajectories, i.e. the functional integral contains a zero mode.
    The contribution of this mode to the functional is not accounted
    for correctly by a straightforward integration over quadratic
    fluctuations. (In fact, we do not yet know how to treat it correctly.)
\end{thebibliography}
\end{document}